\documentclass[pdflatex,sn-nature]{sn-jnl}% Style for submissions to Nature Portfolio journals

\usepackage{times} 
\usepackage{graphicx}
\usepackage{dcolumn}
\usepackage{bm}
\usepackage{color}
\usepackage{xcolor}
\usepackage{hyperref}
\usepackage{enumitem}
\usepackage{cancel}
\usepackage[normalem]{ulem} % for \sout
\usepackage{geometry}

\usepackage{comment}
\hypersetup{colorlinks=true,citecolor=blue,linkcolor=blue, urlcolor=blue}
\newcolumntype{M}[1]{>{\centering\arraybackslash}m{#1}}
\newcolumntype{N}{@{}m{0pt}@{}}
\usepackage{environ}
\usepackage{MnSymbol}
\usepackage{fancyhdr}
\bibpunct{[}{]}{,}{n}{}{}

\usepackage{algorithm}
\usepackage{algpseudocode}
\usepackage[T1]{fontenc}
\usepackage{tikz}

\let\originalleft\left
\let\originalright\right
\renewcommand{\left}{\mathopen{}\mathclose\bgroup\originalleft}
\renewcommand{\right}{\aftergroup\egroup\originalright}

\begin{document}

\title[Article Title]{Exploratory digital alchemy for colloidal crystal discovery}

\author[1]{\fnm{Shih-Kuang (Alex)} \sur{Lee}}\email{shihkual@umich.edu}

\author[2]{\fnm{Sun-Ting} \sur{Tsai}}\email{suntingt@umich.edu}

\author*[1,2,3]{\fnm{Sharon C.} \sur{Glotzer}}\email{sglotzer@umich.edu}

\affil[1]{\orgdiv{Department of Material Science and Engineering}, \orgname{University of Michigan}, \orgaddress{\city{Ann Arbor}, \postcode{48105}, \state{MI}, \country{USA}}}

\affil[2]{\orgdiv{Department of Chemical Engineering}, \orgname{University of Michigan}, \orgaddress{\city{Ann Arbor}, \postcode{48105}, \state{MI}, \country{USA}}}

\affil[3]{\orgdiv{Biointerfaces Institute}, \orgname{University of Michigan}, \orgaddress{\city{Ann Arbor}, \postcode{48105}, \state{MI}, \country{USA}}}

\abstract{
Digital Alchemy (DA), introduced by Van Anders et al., is a statistical mechanics-based generalized thermodynamic ensemble method that employs computer simulations to optimize colloidal particle design. This approach applies the principles of statistical mechanics to predict and tailor particle attributes that lead to desired self-assembled structures or material properties. However, as an inverse design method, its main limitation is that the target structure must be known \textit{a priori}. Therefore, the optimal design from DA does not guarantee the targeted structure is the most or the only stable one. This highlights the importance of forward design with an exploratory scheme for optimizing novel colloid designs, which becomes more suitable in such cases. In this paper, we introduce Exploratory Digital Alchemy (EDA), an enhanced forward design scheme that begins by releasing the constraint of the target crystal from DA, followed by an exploration-oriented bias that has been extensively used in enhanced sampling methods such as metadynamics (MetaD). We demonstrate the utility of EDA through examples involving particles interacting via a two-dimensional Lennard-Jones Gauss potential (LJGP) and a three-dimensional oscillating pair potential (OPP). We applied EDA to study the free energy landscapes given different potential parameters of LJGP at different temperatures. With the exploratory scheme, we've also successfully identified a wide range of OPP potential parameters that stabilize metastable Frank-Kasper phases. Our approach fuses the standard DA framework with metadynamics, which could potentially be useful for studying alchemical reactions in a generalized ensemble.
}

\maketitle

\section*{Introduction}
\label{sec:intro}
Colloidal self-assembly---the spontaneous organization of particles at the micrometer and nanometer scales---underpins advances in materials science~\cite{boles2016self}, nanotechnology~\cite{wang2019controlling}, consumer and beauty~\cite{picken2023sustainable, luengo2023physico}, and drug delivery~\cite{mitchell2021engineering}. The ability to control this process enables the design of materials and products with tailored structures and properties. Digital Alchemy (DA) is an inverse design framework rooted in statistical mechanics that optimizes colloidal particle attributes, such as shape and interparticle interactions, to achieve desired crystal structures~\cite{van2015digital,geng2019engineering,rivera2023inverse}. DA performs simulations in a generalized thermodynamic ensemble, treating particle attributes as thermodynamic variables to target specific self-assembled structures~\cite{van2015digital}. Using methods like Hard Particle Monte Carlo (HPMC)~\cite{van2015digital,geng2019engineering} or Molecular Dynamics (MD)~\cite{zhou2019alchemical,zhou2021inverse}, DA has successfully linked diverse particle designs, including shapes~\cite{D1NR01429C, geng2019engineering}, force field parameters~\cite{zhou2021inverse, zhang2025xchimes}, and patchy interactions~\cite{rivera2023inverse}, to targeted crystal structures. However, as an inverse design method, DA's reliance on \textit{a priori} knowledge of the target structure limits its applicability in exploratory studies, particularly when presented with a novel set of interaction potentials where the target structure is unknown. Additionally, DA only provides nominal optima for particle design based on each simulation. This approach may encounter difficulties due to metastability between particle design parameters and crystal structures: the same set of design parameters can lead to different crystal phases, and conversely, the same crystal phase can be stabilized by different design parameters.

In this paper, we introduce Exploratory Digital Alchemy (EDA), a novel framework that extends the original DA method by combining it with Metadynamics (MetaD)~\cite{parrinello2002wtmetad,hsu2023alchemical}. Unlike DA’s inverse approach, which relies on the constraints of the target structure, EDA releases these constraints and enables forward design by evolving colloidal assembly structures and particle designs across stable phases. EDA harnesses the self-exploration mechanics of MetaD to explore  stable and metastable phases within a generalized thermodynamic  ensemble and identify optimal particle design. That is, EDA allows for a thermodynamically rigorous exploration of particle design space. 

We demonstrate the EDA framework by applying it to explore the design spaces defined by two double-well pairwise potentials: the two-dimensional Lennard-
Jones-Gauss Potential (2D-LJGP) and the three-dimensional Oscillating Pair Potential (3D-OPP) systems, both of which have been shown to assemble a wide range of different crystal structures using a grid search method~\cite{PhysRevLett.98.225505, dshemuchadse2021moving}. We 
demonstrate the simultaneous discovery of crystal structures and their corresponding optimal potential parameter during the evolution of the system by analyzing the free energy surface (FES) of the generalized ensemble. Our method accurately identifies the same crystal structures that have been previously reported. Furthermore, we uncover the metastability among phases in the generalized ensemble, where identical crystal structures can be stabilized by the same optimal potential parameters, or vice versa. By integrating MetaD’s self-exploration feature with the generalized ensemble framework of DA, EDA serves as a powerful tool for discovering colloidal assembly structures and optimizing colloidal particle design, significantly enhancing the original DA framework.

\section*{Results}
\label{sec:results}

\begin{figure}[h]
  \centering
  \includegraphics[width=0.5\textwidth]{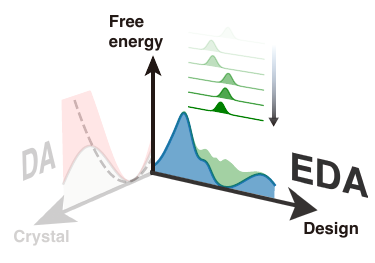}
  \caption
  {\textbf{From Digital Alchemy (DA) to Exploratory Digital Alchemy (EDA)} The EDA modifies the original DA framework, which applies the harmonic constraints (dashed line on the left) on crystal structures, causing the elevation of generalized free energy (red region), and forces the system to equilibrate towards the pre-determined state for inverse design. Instead, EDA combines the generalized ensemble of DA with Metadynamics, which iteratively accumulates Gaussian biases (green region) and enables exploration of the correspondence between crystal structure and particle design.
}\label{fig:abstract}
\end{figure}

\subsection*{Digital Alchemy (DA)}
\label{sec:method_dig_alch}

In DA, particle attributes are treated as thermodynamic variables that fluctuate in a generalized ensemble. These additional thermodynamic variables act as additional degrees of freedom and, because changes in these variables correspond to changes in the particles (e.g., their shape or interparticle interactions), they are referred to as alchemical parameters $\alpha$. So-called "alchemical potentials" -- the conjugate variables to the alchemical parameters -- are obtained by taking the derivative of the generalized free energy with respect to the alchemical parameter while other thermodynamic variables are held fixed. Furthermore, DA applies a harmonic constraint on the system to target a specific crystal structure. Additional details may be found in Refs~\cite{van2015digital,geng2019engineering,rivera2023inverse, D1NR01429C,zhou2019alchemical,zhou2021inverse, zhang2025xchimes} and the Methods section. 

\subsection*{Exploratory Digital Alchemy (EDA)}
\label{sec:method_metad}

Instead of imposing constraints to bias the sampling towards a specific crystal structure, we propose an exploratory version of Digital Alchemy (EDA) that samples \textit{all microstates} (including all thermodynamically stable and metastable states) defined by $\{X,\alpha\}$ in the generalized ensemble. To perform EDA, we apply an $\alpha$-dependent bias potential to facilitate and promote the exploration of different crystal structures, resulting in the biased partition function:
\begin{align}
    \mathcal{Z} = \sum_{X,\alpha} e^{-\beta [ U_\alpha(X) + V(\mathbf{s}) ]}
    \label{eq:Z_DA_MetaD}
\end{align}
where $X$ denotes a microstate of the system, $\beta=1/k_BT$ is the inverse of the product of the Boltzmann constant $k_B$ and temperature $T$, and $V(\mathbf{s})$ is the bias potential constructed as a functional of a set of collective variables $\mathbf{s}=\mathbf{s}(X,\alpha)$.  In this work, we apply a history-dependent bias potential that has been extensively applied in Well-Tempered Metadynamics (MetaD) and is constructed iteratively, as described in~\cite{parrinello2002wtmetad,barducci2008well,valsson2016enhancing,bussi2020using}. We provide a detailed description in the Methods section. 

\noindent With the help of MetaD, we are able to explore the generalized free energy $\varphi$,  defined as
\begin{align}
    \varphi = -\frac{1}{\beta}\ln\sum_{X,\alpha} e^{-\beta U_{\alpha}(X)}
\end{align}
In practice, we usually consider the free energy surface as a function of the collective variables, sometimes called potential of mean force (PMF), $\Phi$
\begin{equation}
\begin{split}
    \Phi(\alpha, w) =  -&\frac{1}{\beta} \ln  \sum_{X',\alpha'} e^{-\beta U_{\alpha'}(X')} \delta(w-w(X'))\delta(\alpha-\alpha') 
\end{split}
\end{equation}
where $w=w(X)$ is the collective variable, often defined to identify the metastable states sampled during the simulation. In this paper, we will use the term free energy surface (FES) and PMF interchangeably. Additionally, we will apply our method to sample the Helmholtz free energy, $F$, over all possible $\alpha$
\begin{align}
    F(\alpha) = -\frac{1}{\beta}\ln\int e^{-\beta \Phi(\alpha, w)} dw
\end{align}

In the following sections, we demonstrate the utility of EDA in broadening the search for optimal particle interactions that facilitate self-assembly of many thermodynamically stable and metastable structures in a single simulation. As examples, we apply EDA to systems of particles interacting via two types of double-well potentials: ($\textrm{i}$) the two-dimensional Lennard-Jones-Gauss Potential (LJGP)~\cite{rechtsman2005optimized, PhysRevLett.98.225505}  and ($\textrm{ii}$) the three-dimensional Oscillating Pair Potential (OPP)~\cite{mihalkovivc2012empirical,elenius2009structural}. Systems of particles interacting via these two pair potentials self-assemble into a wide range of different crystal structures and thus are well-suited to evaluate the exploratory power of EDA. Additionally, we perform the Minkowski Structure Metric (MSM) order parameter calculation, $w_l$, and other structural analyses, including diffraction pattern analysis and bond-orientational order calculations, to help identify the crystal structures discovered by EDA. Details can be found in the Methods section.

\subsection*{Lennard-Jones Gauss Potential (LJGP)}
\label{sec:res_ljg}
\begin{figure*}[th!]
  \centering
  \includegraphics{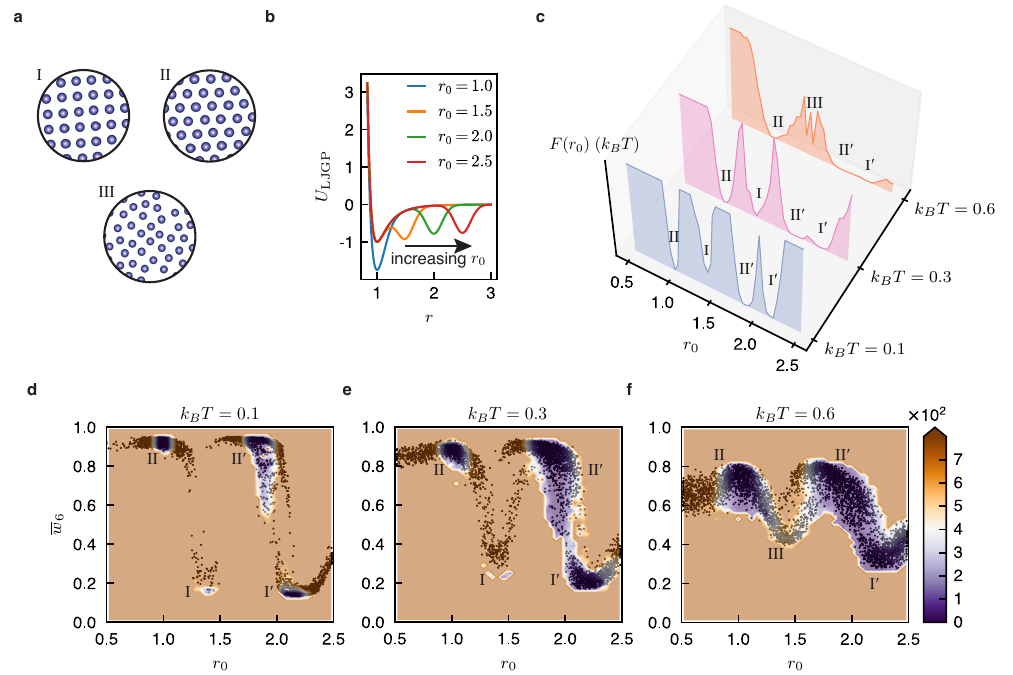}
  \caption
  {\textbf{Two-dimensional Lennard-Jones-Gauss Potential (LJGP) system.} \textbf{a} Thermodynamic phases discovered by the EDA method, including: \textbf{I} square lattice, \textbf{II} hexagonal lattice, and \textbf{III} disordered phase. \textbf{I}' and  \textbf{II}' represent the same crystal structure with different lattice constants \textbf{b} The LJGP energy profile at different $r_0$. \textbf{c} Helmholtz free energy surface $F$ (FES) as a function of $r_0$ and temperature. \textbf{d}, \textbf{e}, and \textbf{f} The points of collective variables ($r_0$, $\overline{w}_6$) sampled by the EDA method and the corresponding FES at various temperatures. The minima on the FES are indicated by the corresponding phases in \textbf{a}. The colorbar indicates the FES value $\Phi(r_0, \overline{w}_6)$ in units of $k_BT$.}
\label{fig:ljgp_freeE_alpha}
\end{figure*}

As our first test case, the Lennard-Jones Gauss Potential (LJGP) is a pairwise double-well potential constructed by adding a negative Gaussian function to the scaled Lennard-Jones potential~\cite{rechtsman2005optimized, PhysRevLett.98.225505}. The Lennard-Jones component in the LJGP has an energy minimum at $r=d=1$. The LJGP has a parameter $r_0$ that controls the position of the second potential energy well as shown in Fig.~\ref{fig:ljgp_freeE_alpha} \textbf{b}, and determines the resulting crystal structure. Henceforth, we use the $r_0$ as our alchemical parameter. We performed EDA simulations with $400$ particles in 2D interacting via the LJGP with a fixed volume of $60 d^2$ at temperatures $k_BT=0.1$, $0.3$ and $0.6$. For the purposes of statistical analysis, we ran five independent replicas at each temperature.

The EDA simulations were implemented with $r_0$ as the fluctuating thermodynamic alchemical parameter. The bias potential is constructed as a function of $r_0$ following the approach outlined in the Methods section. For each replica, we prepared the simulations by first setting $r_0=1.5$ as an initial guess. We then cooled the system from $k_BT=1.0$ to the target temperatures using 100000 MC sweeps. We calculated $\overline{w}_6$ to identify the thermodynamic phases arising during the simulations and then constructed the unbiased Boltzmann distribution and FES at each target temperature as a function of $(\overline{w}_6, r_0)$, where $\overline{w}_{\ell}$ is defined as the system average of $w_{\ell}$ calculated for each individual particle. The potential energy profiles of different $r_0$ are also illustrated in FIG.~\ref{fig:ljgp_freeE_alpha} \textbf{b}.

Our findings for the LJGP system are reported in FIG.~\ref{fig:ljgp_freeE_alpha}. At $k_BT=0.1$, we observed two stable, ordered structures: a square lattice (FIG.~\ref{fig:ljgp_freeE_alpha} \textbf{a} $\textrm{I}$) and a hexagonal lattice (FIG.~\ref{fig:ljgp_freeE_alpha} \textbf{a} $\textrm{II}$). The square lattice is stable at two different ranges of $r_0$: $r_0 = 1.3$ to $1.5$ and $r_0 = 2.0$ to $2.5$. These stability ranges are labeled as $\textrm{I}$ and $\textrm{I}'$ and correspond to square lattices with different lattice constants. The hexagonal lattice is stable within two different ranges $r_0$: $r_0 = 0.5$ to $1.3$ and $r_0 = 1.5$ to $2.0$, which are labeled as $\textrm{II}$ and $\textrm{II}'$ and correspond to hexagonal lattices with different lattice constants. FIG.~\ref{fig:ljgp_freeE_alpha} \textbf{c} shows the FES profile $F(r_0)$ by integrating out $\overline{w}_6$. We note that the $F(r_0)$ in FIG.~\ref{fig:ljgp_freeE_alpha} \textbf{c}  is smoothed via the Kernel Density Approximation with Gaussian kernels~\cite{scott2011multivariate, silverman2018density} to first smooth the reweighted probability density function and calculate the $F(r_0)$. At $k_BT=0.1$, the free energy basin of square lattice $\textrm{I}$ is isolated by high barriers, and its free energy minimum is relatively higher than the other phases. 
As $k_BT$ is increased above $0.3$, the square lattice $\textrm{I}$ becomes more stable, and the free energy minimum becomes lower, and the barriers that separate it from the other phase become lower, as shown by the pink FES in FIG. \ref{fig:ljgp_freeE_alpha} \textbf{c}. Same results for $k_BT=0.1$ and $0.3$ can also be seen from FIG.~\ref{fig:ljgp_freeE_alpha} \textbf{d} and \textbf{e} that the square lattice $\textrm{I}$ is less stable than the other phases: the free energy basin of $\textrm{I}$ is substantially higher than the other phases. However, although at $k_BT=0.6$ there is still a visible free energy minimum, it slightly narrows and splits, and the system becomes disordered (indicated by $\textrm{III}$), as shown by the orange FES in FIG.~\ref{fig:ljgp_freeE_alpha} \textbf{b}. Additionally, we observe a corresponding upward shift in the energy basin location, indicated by $\textrm{III}$ in \textbf{f}, which used to be $\textrm{I}$ in \textbf{e}. Additionally, the free energy barrier between the square lattice $\textrm{I}'$ and the hexagonal lattice $\textrm{II}'$ disappears at that temperature, at which we expect that the transitions become easier at higher $ k_BT$. We can also see the disappearance of energy barrier in the 2D plot of the FES (FIG.~\ref{fig:ljgp_freeE_alpha} \textbf{f} that the barrier between $\textrm{I}'$ and $\textrm{II}'$ nearly vanishes as $k_BT$ is increased to $0.6$.

% \begin{figure*}[!th]
%   \centering
%   \includegraphics[width=\textwidth]{fig3.pdf}
%   \caption
%   {\textbf{Equilibrium phases obtained from cooling simulations for the Lennard-Jones-Gauss Potential system with $r_0=1.8$ to $2.1$.}  The four snapshots indicated by the grey box in the middle are the $r_0$ values that we intend to confirm that the rhombic lattice is unstable at a finite temperature.
% }\label{fig:ljgp_cooling}
% \end{figure*}

We compare our results at $k_BT=0.1$ to the zero temperature phase diagram reported by Engel et. al.~\cite{PhysRevLett.98.225505}, in which an exhaustive grid search of LJGP potential parameters $(\epsilon_G, r_0)$ was performed. The stable crystal structures identified at $k_BT=0.1$ and shown in FIG.~\ref{fig:ljgp_freeE_alpha} mostly align with Engel's results, with the exception of the rhombic lattice, which was reported to be stable within a narrow range of $r_0 = 2.0$ to $2.04$ but is not observed with our method. We attribute this discrepancy to the fact that our simulations were performed at a finite temperature. We confirm using standard MC simulations, starting from the same initial configuration as in the EDA simulation and cooling the system, starting from $k_BT=1.0$ down to $k_BT=0.1$ using 2 million steps. We initiated the simulations from multiple points within the range of $r_0 = 2.0$ to $2.04$, and we only observe the hexagonal crystal as the dominant equilibrium phase. 

% We demonstrate the snapshot of those equilibrium phases and their corresponding diffraction pattern in FIG.~\ref{fig:ljgp_cooling}.

\subsection*{Oscillating Pair Potential (OPP)}
\label{sec:res_opp}

\begin{figure*}[!th]
  \centering
  \includegraphics{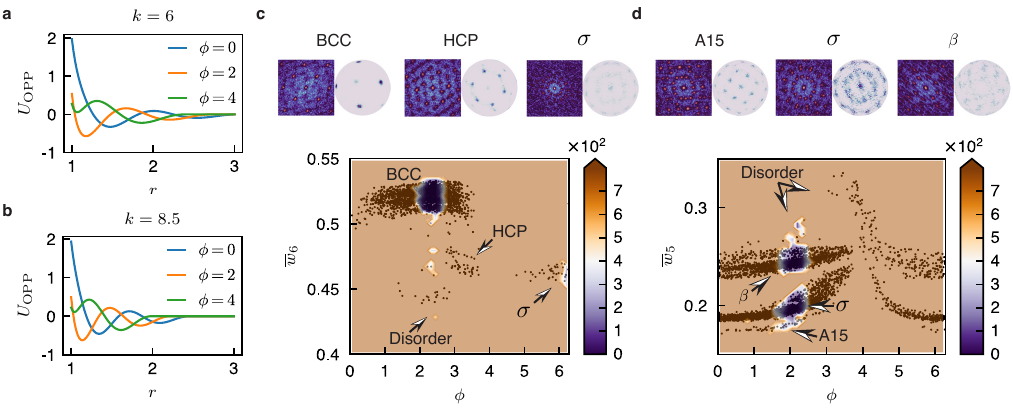}  % 6.75 inches
  \caption
  {\textbf{Three-dimensional Oscillating Pair Potential (OPP) system.} \textbf{a} and \textbf{b} The OPP energy profile for $k=6$ and $8.5$ respectively, at different $\phi$. \textbf{c} and \textbf{d} The points of collective variables ($\phi$, $\overline{w}_l$) sampled by the EDA method and the corresponding free energy surface (FES), for $k=6$ and $8.5$, respectively. In each subfigure of \textbf{c} and \textbf{d}, the top row plots the diffraction pattern and bond-orientational order diagram for the three crystals discovered, and the FES minima at the bottom row are indicated by the corresponding phases. The colorbars indicate the FES value $\Phi(\phi, \overline{w}_l)$ in units of $k_BT$.
}\label{fig:opp_hist}
\end{figure*}

In this subsection, we present our findings of the double-well pair potential, OPP, in three-dimensional systems. The OPP was originally proposed to approximate electronic interaction of intermetallic systems and is based on the form of the Mihalkovi\v c Henley interaction potential~\cite{mihalkovivc2012empirical,elenius2009structural}. We show the potential energy profile in FIG.~\ref{fig:opp_hist} \textbf{a} and \textbf{b}. The potential energy profile is controlled by the two parameters $k$ and $\phi$ which are the wavenumber and phase shift, respectively, of a damped oscillation in the profile. Here, we use the $\phi$ as our alchemical parameter.

We performed EDA simulations with point particles interacting via the OPP and let the potential parameter $\phi$ fluctuate while the wave number $k$ is kept fixed. The bias potential is constructed as a function of $\phi$, which we use as a collective variable defined within the interval $(0, 2\pi)$. We prepared the simulations at two different wave numbers $k=6$ and $k=8.5$ for comparison. For $k=6$, we performed five independent simulations with $1000$ particles each, and for $k=8.5$, we performed five independent simulations with $810$ particles each. Each simulation was initialized with different $\phi$ values, set as integers from $1$ to $5$. We also prepared the systems by cooling them from $k_BT=0.5$ to $k_BT=0.1$ over 5 million MC sweeps. The sampled points via EDA and the corresponding FES are constructed as a function of $(\overline{w}_6, \phi)$ for $k=6$ and as a function of $(\overline{w}_5,\phi)$ for $k=8.5$, and shown in FIG.~\ref{fig:opp_hist} \textbf{c} and \textbf{d}.

\begin{figure*}[!th]
  \centering
  \includegraphics{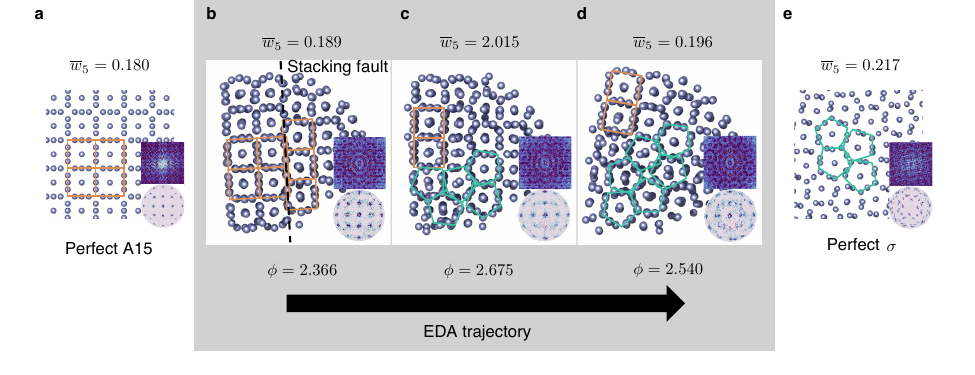}  % 6.5 inches
  \caption
  {\textbf{Transition between A15 and $\sigma$ phases at $k=8.5$.} \textbf{a} The perfect A15 phase. \textbf{e} The perfect $\sigma$ phase. \textbf{b} The beginning, \textbf{c} intermediate, and \textbf{d} final stage of the transition. In \textbf{b}, the stacking fault is indicated by the black dashed line. The special motifs of A15 and the $\sigma$ phase are highlighted by the brown square and green pentagons, respectively.
}\label{fig:opp_a15_to_sigma}
\end{figure*}

We first present the results obtained with wave number $k=6$, shown in FIG~\ref{fig:opp_hist} \textbf{a} and \textbf{c}. Using $\overline{w}_6$ as the collective variable, we identified four distinct thermodynamic phases: body-centered cubic (BCC), hexagonal close-packed (HCP), a disordered phase, and $\sigma$ phase, a Frank-Kasper phase with complex unit cell. Of the phases identified, the BCC crystal appears to be the most stable and possesses the lowest free energy located at $\overline{w}_6 \approx 0.52$. The FES indicates that the optimal value of the alchemical parameter $r_0$ for stabilizing BCC is near $2.3$. As $r_0$ approaches $3$, we observed several transitions from BCC to HCP ($\overline{w}_6 \approx 0.475$); however, the missing free energy minimum where HCP is observed shows that HCP is significantly less stable than BCC. The $\sigma$ phase ($\overline{w}_6 \approx 0.46$), though still less stable than BCC, is stable with a noticeable free energy minimum when $r_0$ is near $6.0$. 

Next, we apply EDA to the OPP system with wave number $k=8.5$. As shown in FIG.~\ref{fig:opp_hist} \textbf{b} and \textbf{d}, we observed four phases at the values of the collective variable indicated in parentheses: The A15 phase ($\overline{w}_5 \approx 0.18$), $\sigma$ phase ($\overline{w}_5 \approx 0.19$), $\beta$-manganese crystal ($\overline{w}_5 \approx 0.24$), and a disordered phase. The three crystal structures are stable when the alchemical parameter $\phi \approx 2.0$; each is associated with a corresponding free energy minimum. Due to the lower free energy barrier between A15 and $\sigma$ phases, we also observed several transitions back and forth between them during the simulations, with the $\sigma$ phase being more stable. We illustrate the transition between A15 and $\sigma$ phases in FIG.~\ref{fig:opp_a15_to_sigma}. The transition begins with the defect formation, such as a stacking fault in the A15 phase, which creates instability, as shown in FIG.~\ref{fig:opp_a15_to_sigma} \textbf{b}. Subsequently, when the $\phi$ value evolves to approximately $2.5$, the incommensurable A15 phase and the interaction introduce local lattice distortion, causing the transformation from the square brown motifs of A15 to the pentagon-like green motifs of $\sigma$ phase. This leads to the transformation of the A15 phase into the $\sigma$ phase near the defects, as depicted in FIG.~\ref{fig:opp_a15_to_sigma} \textbf{c}. Finally, in FIG.~\ref{fig:opp_a15_to_sigma} \textbf{d}, the $\sigma$ phase starts to expand into the surrounding structure. For reference, we also demonstrate the perfect A15 and $\sigma$ phases in FIG.~\ref{fig:opp_a15_to_sigma} \textbf{a} and \textbf{e}. The relatively low free energy barrier between the $\sigma$ and A15 phases may be due to their similar local structural motifs. As reported by Dshemuchadse et al.~\cite{dshemuchadse2021moving}, both phases have a high coordination number of 13.5 and contain similar motifs in their unit cell, with one particle surrounded by two alternating hexagons formed by the other 12 particles. On the other hand, we observed no transitions between $\beta$-manganese and the other two crystals, which is consistent with the significantly high free energy barrier  ($>700k_BT$!) around the $\beta$-manganese basin.  Although sampling transition states between crystal phases and estimating the corresponding free energy barriers are beyond the scope of this paper and are left as future work, we note that the poor sampling of transitions between crystal phases can be improved by constructing the bias potential as a function of the pair $(\overline{w}_5, \phi)$.

\begin{figure*}[!th]
  \centering
  \includegraphics{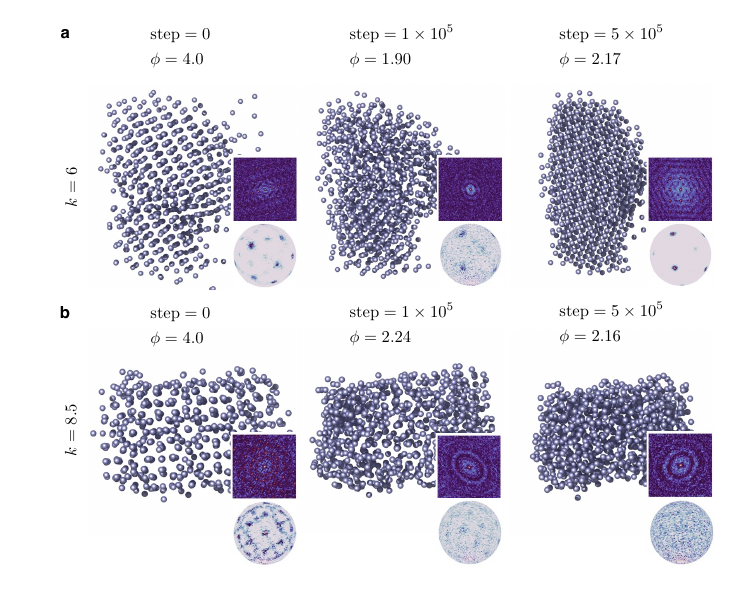}  % 5 inches
  \caption
  {\textbf{Stability test for Oscillating Pair Potential system.} \textbf{a} The evolution of the unbiased EDA simulation snapshots at $k=6$, starting from the equilibrium $hP2$ phase at $k_BT=0.1$ and $\phi=4$. \textbf{b} The evolution of the unbiased EDA simulation snapshots at $k=8.5$, starting from the equilibrium $cP54$ phase at $k_BT=0.1$ and $\phi=4$. 
}\label{fig:opp_eda_nobias}
\end{figure*}

We can compare our findings to those reported in a detailed mapping of the OPP phase diagram for different values of $(k, \phi)$ reported in Dshemuchadse et al.~\cite{dshemuchadse2021moving}. The EDA simulations discovered crystals at consistent values of $\phi$ for the two values of $k$ we studied. However, instead of the $hP2$ (compressed hcp/$hP2$ Mg) and $cP54$ (clathrate-I) phases reported by Dshemuchadse et al.~\cite{dshemuchadse2021moving}, EDA found a disordered phase at the same location of $\phi=4$. This discrepancy can be explained by the instability of the complex crystals for the values of $\phi$ and $k$ we explored, as Dshemuchadse et al.~\cite{dshemuchadse2021moving} found $hP2$ and $cP54$ to be stable only within a very narrow $\phi$ regime. In FIG.~\ref{fig:opp_eda_nobias}, We perform stability tests to confirm by setting up both crystals, running a quick EDA simulation, and setting the bias potential to zero, i.e., allowing $\phi$ to fluctuate freely, and found that neither crystal is stable in our exploratory regions.

For $k=8.5$, the three main crystals (A15, $\sigma$, and $\beta$-manganese phases) discovered by EDA exhibit comparable stability, each with nearly the same $\phi\approx 2$. In other words, a single particle design with $\phi\approx 2$ can stabilize multiple phases simultaneously. In fact, the free energy plot in FIG.~\ref{fig:opp_hist} \textbf{d} demonstrates that the sampled points for these three crystals distribute as three nearly horizontal strips along the $\phi$ direction, indicating that all three phases can remain metastable across a broad range of $\phi$. This result highlights the advantage of EDA over the DA method, as EDA provides a more comprehensive perspective on metastability within the generalized ensemble and identifies all metastable phases for each particle design. To confirm the stability of crystals obtained through the EDA simulations, we conducted a stability test using the $k=8.5$ OPP system, initialized from sampled snapshots of three crystals at $\phi=2$ in the middle of the EDA simulation. We then performed a standard NVT MC simulation. Each system remained stable after running $5$ million MC simulation steps. This suggests that all the sampled crystals are either thermodynamically stable or metastable at $\phi=2$.

\section*{Discussion}
\label{sec:conclusion}

In summary, we proposed and demonstrated a new approach---Exploratory Digital Alchemy (EDA)---to discovering thermodynamically stable and metastable crystal structures within design space defined by pair potentials and their alchemical parameters. To improve the original Digital Alchemy (DA) framework by incorporating a self-exploration feature, we employed a history-dependent bias potential commonly used in the well-known metadynamics method. We applied EDA to systems of particles interacting via isotropic multi-well pair potentials: a two-dimensional system with the Lennard-Jones Gauss Potential (LJGP) and a three-dimensional system with the Oscillating Pair Potential (OPP). For LJGP, we successfully reproduced the stable phases previously reported at low temperature. In addition, we observed transitions between metastable phases at higher temperatures for the first time. For OPP, we tested EDA with two different wave numbers $k$. At a lower $k$, EDA produced three metastable crystal phases with two simple lattice structures and one Frank-Kasper phase. At higher $k$, EDA produced three metastable complex crystals, including two Frank-Kasper phases. Besides, our EDA method reveals the metastability in the generalized ensemble for OPP system, as the same particle design can stabilize multiple crystal structures. Our work tackles the sampling problem in design space by introducing an enhanced sampling method to digital alchemy. The combination of the two methods also enables quantitative analysis of the free energy surface as a function of the alchemical degree of freedom.

\section*{Methods}
\label{sec:method}
\subsection*{Details about Digital Alchemy (DA)}
Consider a system of isotropic point particles interacting with a pair potential $U_\alpha(r)$, in which there is one alchemical parameter $\alpha$ (this can be easily generalized to multiple alchemical parameters).  The goal of DA is to find the value of $\alpha$ that minimizes the free energy of the system of point particles in the target crystal structure, in the hope that by using this optimized alchemical parameter, the system will self-assemble into the target structure from an initially disordered state.

To target a desired crystal structure, the biased partition function of the generalized ensemble $\mathcal{Z}$, i.e., biased NVT$\mu_\alpha$ ensemble, is defined as~\cite{van2015digital, metropolis1953equation}
\begin{align}
    \mathcal{Z} = \sum_{X,\alpha} e^{-\beta [ U_\alpha(X) -\mu_\alpha N\alpha - \lambda\Lambda ]}
    \label{eq:Z_DA}
\end{align}
where $X$ denotes a microstate of the system, $\beta=1/k_BT$ is the inverse of the product of the Boltzmann constant $k_B$ and temperature $T$, and $N$ is the number of particles. The alchemical potential, $\mu_\alpha$, is the thermodynamic conjugate of $\alpha$. The Lagrange multiplier $\lambda>0$ is associated with a constraint $\Lambda$ that drives the system toward the target crystal structure. In a conventional DA simulation, one applies a spring constant $K=\lambda\beta$ and constraint $\Lambda=\|\mathbf{r}-\mathbf{r}^{(0)}\|^2$, where $\mathbf{r}$ and $\mathbf{r}^{(0)}$ are the particle positions in the system and the reference position of the target crystal, respectively. The $e^{\beta\lambda\Lambda}$ term in Eq.~\ref{eq:Z_DA} can then be seen as constraining the system to sample the variants of an Einstein crystal~\cite{van2015digital}. 

To sample the partition function $\mathcal{Z}$ defined in Eq.~\ref{eq:Z_DA} and find the optimal $\alpha$, we apply the Metropolis algorithm as follows for determining the acceptance ratio of a move from alchemical parameter $\alpha_1$ to $\alpha_2$ in a Monte Carlo simulation:
\begin{align}
    a_{\alpha_1\to\alpha_2} = \min\left\{ 1, \frac{e^{-\beta U_{\alpha_2}(X)}}{e^{-\beta U_{\alpha_1}(X)}} \right\}
    \label{eq:metropolis_DA}
\end{align}
The simulation is performed with fluctuating $\alpha$ at constant $\mu_\alpha=0$~\cite{van2015digital}. We note that while the MD sampling of the partition function $\mathcal{Z}$ is possible, as demonstrated in the Refs~\cite{zhou2019alchemical,zhou2021inverse} where the alchemostat is implemented following the Nos\'e-Hoover's formulation~\cite{martyna1994constant} to control the  $\mu_\alpha$, we choose to perform MC sampling with a more straightforward implementation. This choice simplifies the development process of a new method, such as the EDA in this paper.

\subsection*{Combine Well-Tempered Metadynamics (MetaD) with DA}
MetaD~\cite{parrinello2002wtmetad,barducci2008well,valsson2016enhancing,bussi2020using} is a bias simulation technique that iteratively accumulates Gaussian bias during the course of simulation as: 
\begin{align}
    \begin{aligned}
        V(\mathbf{s}, t+1) &= V(\mathbf{s}, t) + W_t\exp\left[-\frac{\|\mathbf{s}-\mathbf{s}_t\|^2}{2\sigma^2}\right] \\
        W_t &= \omega e^{-\beta V(\mathbf{s}_t) / (\gamma-1)}
        \label{eq:bias_update}
    \end{aligned}
\end{align}
where $V(\mathbf{s}, 0)=0$ and the update of the bias potential is performed periodically every $\tau$ timestep. The $\mathbf{s}_t$ are the set of collective variables calculated from the system evaluated at timestep $t$, $\omega$ is the Gaussian height, $\sigma$ controls the width of the Gaussian bias, and $\gamma$ is a user-defined parameter controlling the adaptation of the Gaussian height. Empirically, the DA method sometimes suffers from sensitivity to the initial alchemical parameters it begins with~\cite{zhang2025xchimes}. This is primarily due to high barriers present in alchemical space that isolate local minima of $\alpha$, resulting in a non-ergodicity problem. Therefore, to maximize the exploration capability of the bias potential, we will set $\mathbf{s}=\alpha$, where $\alpha$ are the alchemical parameters that parameterize the interacting potentials in the system. Eq.~\ref{eq:metropolis_DA} then becomes
\begin{align}
    a_{\alpha_1\to\alpha_2} = \min\left\{ 1, e^{\beta(\Delta U+\Delta V)} \right\}
    \label{eq:metropolis_DA_MetaD}
\end{align}
where $\Delta U=U_{\alpha_2}(X)-U_{\alpha_1}(X)$ and $\Delta V=V(\alpha_2,t+1)-V(\alpha_1,t)$. Following the standard approach of free energy estimation used in metadynamics~\cite{valsson2016enhancing} we can also reweight the sampled collective variable $\xi=\xi(X,\alpha)$ and estimate the unbiased Boltzmann distribution using the time-dependent bias potential $V(\mathbf{s},t)$ constructed on-the-fly~\cite{tiwary2015time}. Note that the $\xi$ for analyzing the statistics can be different from the $\mathbf{s}$ which is used to define the bias potential:
\begin{align}
    \langle \xi(X,\alpha) \rangle
    &=\langle \xi(X,\alpha) e^{-\beta [V(\mathbf{s}(t),t)-c(t)]}\rangle_{V} \\
    e^{-\beta c(t)}
    &\equiv\frac{
    \int e^{-\frac{\gamma}{\gamma-1}\beta V(\mathbf{s},t)}d\mathbf{s}
    }{
    \int e^{-\frac{1}{\gamma-1}\beta V(\mathbf{s},t)}d\mathbf{s}
    }
    \label{eq:reweighting_factor}
\end{align}
where $\langle\cdot\rangle_V$ represents the average over the biased ensemble.

We summarize the MetaD parameters used in this paper in the Table~\ref{tab:MetaD_param}:

\begin{table}[ht]
\renewcommand{\arraystretch}{1.3} % Adjust row spacing
\setlength{\tabcolsep}{2.8mm} % Adjust column spacing
\begin{tabular}{ccccc}
\toprule
\multicolumn{5}{c}{Metadynamics parameters} \\
\hline
Pair potential & $\omega\; (k_BT)$ & $\sigma$ & $\tau$ & $\gamma$ \\ \hline
LJGP, $k_BT=0.1$        & 300 & 0.01 & 1000 & 2000 \\ \hline
LJGP, $k_BT=0.3$        & 100 & 0.01 & 1000 & 2000 \\ \hline
LJGP, $k_BT=0.6$        & 50 & 0.01 & 1000 & 2000 \\ \hline
OPP, $k=6$   & 1000 & 0.05 & 10000 & 5000 \\ \hline
OPP, $k=8.5$   & 3000 & 0.05 & 10000 & 8000 \\
\toprule
\end{tabular}
\caption{Metadynamics parameters used in this study.}
\label{tab:MetaD_param}
\end{table}

\subsection*{Simulation Details}
\label{sec:simulation software}
The two pair potentials used in this paper in the Results section are shown here. The LJGP~\cite{rechtsman2005optimized, PhysRevLett.98.225505} is defined as:
\begin{align}
    U_{\rm LJGP}(r)
    =\left[
    \left(\frac{1}{r}\right)^{12}-2\left(\frac{1}{r}\right)^6
    \right]
    -\epsilon_{\rm G}\exp{\left(-\frac{(r-r_0)^2}{2\sigma_{\rm G}^2}\right)}
\end{align}
Here we use $\epsilon_G=0.75$, $\sigma_G^2=0.02$, $r$ is the interparticle distance, and $r_0$ is the center of the Gaussian function and controls the position of the second potential energy well. We use the $r_0$ as the alchemical parameter for the LJGP system.

For the OPP~\cite{mihalkovivc2012empirical,elenius2009structural}, the potential is defined as:

\begin{align}
    U_{\rm OPP}(r)=\frac{1}{r^{15}} + \frac{\cos{(k(r-1)+\phi)}}{r^3}
\end{align}
where $k$ and $\phi$ are the wavenumber and phase shift, respectively, of a damped oscillation. The oscillation is truncated and shifted to zero at the maximum following the second attractive well.  We use the $\phi$ as the alchemical parameter for the OPP system.

We performed the Monte Carlo (MC) simulations to sample the NVT and NVT$\mu_\alpha$ ensembles using the HPMC integrator implemented in HOOMD-blue~\cite{anderson2020hoomd} version 5.0, which can be applied to particle systems interacting via a pair potential. For both systems, EDA simulations are performed by initializing a random system with a fixed initial guess of the alchemical parameter, followed by a cooling simulation to a low enough temperature; then, we start the EDA simulations that perturb the alchemical parameter and iteratively accumulate the bias potential as described by Eq.~\ref{eq:bias_update}. The alchemical parameter MC move and the bias potential calculation are implemented using the custom updater module of HOOMD-blue. We perform the Minkowski Structure Metric (MSM), $w_l$, order parameter calculation and other structural analyses, including diffraction pattern analysis and bond-orientational order calculations, using freud~\cite{freud2020}. The simulation snapshots are rendered using Ovito~\cite{stukowski2009visualization}.

\subsection*{Minkowski Structure Metric (MSM)}
\label{sec:method_cv}

Because the size, symmetry and lattice constant of the sampled crystal structures tend to constantly fluctuate as a result of $\alpha$ fluctuating, we need a robust collective variable to explore crystal structures that does not rely on a radial cutoff function. Here, we used the MSMc~\cite{mickel2013shortcomings}, where the neighboring particles are weighted and determined based on the Voronoi tessellation, as the order parameter for our collective variable. The MSM $w_{\ell,i}$ of particle $i$ is considered an  improved version of the Steinhardt OP~\cite{lechner2008accurate} and is defined as:
\begin{align}
w_{\ell, i} &= \sqrt{\frac{4 \pi}{2\ell+1}\sum_{m=-\ell}^{\ell}w_{\ell m, i}w_{\ell m, i}^{*}} \\
w_{\ell m, i} &= \frac{1}{\sum_{j=1}^{N_b}A_{ij}}\sum_{j=1}^{N_b}A_{ij} Y_{\ell m} (\boldsymbol{r}_{ij})
\end{align}
where $A_{ij}$ is the surface area of the Voronoi cell facet separating the two neighboring particles $i$ and $j$, $Y_{\ell m}$ is a spherical harmonic, and $\boldsymbol{r}_{ij}$ is the relative position vector between neighboring particles. Note that for a two-dimensional system, the calculation of MSM simplifies as the polar angle, $\phi_{ij}$, alone is sufficient to define the bond orientation:
\begin{align}
w_{\ell, i} &=  \left| \frac{1}{\sum_{j=1}^{N_b}A_{ij}}\sum_{j=1}^{N_b}A_{ij} e^{i\ell \phi_{ij}} \right|
\end{align}
Depending on the crystal structures sampled, other local OPs may be useful as CVs, such as the recently proposed Point Group Order Parameter~\cite{fijan2025quantifying} . Continuous CVs are also possible~\cite{zhao2026hybrid}.

\section*{Funding}
This research was supported by the National Science Foundation, Division of Materials Research under a Computational and Data-Enabled Science and Engineering (CDS\&E) Award number DMR 2302470, and by a Vannevar Bush Faculty Fellowship sponsored by the Department of the Navy, Office of Naval Research under ONR award number N00014-22-1-2821. Computational resources and services were provided by Advanced Research Computing at the University of Michigan, Ann Arbor.

\section*{Conflict of interest/Competing interests}

The authors declare no competing interests.

\section*{Data availability}

The trajectory data used to generate the results in this paper can be accessed through the University of Michigan’s Deep Blue Data repository via the DOI: \url{https://doi.org/10.7302/34xw-w788}.

\section*{Author contribution}
S.C.G. supervised the research. S.K.L., S.T.T., and S.C.G. designed the research. S.K.L. designed and performed the Exploratory Digital Alchemy simulations. S.K.L. and S.T.T. provided the simulation data analysis and interpretation. All authors contributed to the writing. All authors read and approved the final manuscript.

\bibliography{references.bib}

\end{document}